\RequirePackage{fix-cm}
\documentclass[smallextended]{svjour3}       
\smartqed  
\usepackage{graphicx}
\usepackage{algorithmic}
\usepackage{comment}
 \usepackage{natbib}
\begin{document}

\title{Using binary decision diagrams for constraint handling in combinatorial interaction testing
}


\author{Tatsuhiro Tsuchiya}


\institute{T. Tsuchiya \at
	          Graduate School of Information Science and Technology, Osaka University \\
              1-5 Yamadaoka, Suita-City, Osaka 565-0871, Japan \\
              Tel.: +81-6-68794535\\
              Fax: +81-6-68794539\\
              \email{t-tutiya@ist.osaka-u.ac.jp}           
}

\date{Received: date / Accepted: date}

\maketitle

\begin{abstract}
Constraints among test parameters often have substantial effects on the performance of test case generation for combinatorial interaction testing. This paper investigates the effectiveness of the use of Binary Decision Diagrams (BDDs) for constraint handling. BDDs are a data structure used to represent and manipulate Boolean functions. The core role of a constraint handler is to perform a check to determine if a partial test case with unspecified parameter values satisfies the constraints. In the course of generating a test suite, this check is executed a number of times; thus the efficiency of the check significantly affects the overall time required for test case generation. In the paper, we study two different approaches. The first approach performs this check by computing the logical AND of Boolean functions that represent all constraint-satisfying full test cases and a given partial test case.  The second approach uses a new technique to construct a BDD that represents all constraint-satisfying partial test cases. With this BDD, the check can be performed by simply traversing the BDD from the root to a sink. We developed a program that incorporates both approaches into IPOG, a well-known test case generation algorithm.
Using this program, we empirically evaluate the performance of these BDD-based constraint handling approaches
using a total of 62 problem instances.  
In the evaluation, the two approaches are compared with three different constraint handling approaches, namely, 
those based on Boolean satisfiability (SAT) solving, Minimum Forbidden Tuples (MFTs), and Constraint 
Satisfiction Problem (CSP) solving. 
The results of the evaluation show that the two BDD-based approaches usually outperform the other 
constraint handling techniques and that the BDD-based approach using the new technique exhibits best performance.
\end{abstract}

\section{Introduction}\label{sec:intro}

In testing of real-world systems, executing all possible test cases is impractical,  
as the number of the test cases is usually prohibitively large.  
This motivates the development of cost reduction strategies.  
\emph{Combinatorial Interaction Testing (CIT)}~\citep{Grindal2005,Nie:2011,intro}  
is one of these testing strategies. 
CIT requires the tester to test all parameter interactions of interest, 
rather than all possible test cases.   
The most basic and well-practiced form of CIT is \emph{$t$-wise testing}. 
This strategy requires testing all value combinations among any $t$ parameters. 
Previous studies show that small $t$ is often sufficient to detect a 
significant percentage of defects~\citep{998624}.  
Hence this can greatly reduce the number of test cases to execute. 
The value $t$ is usually referred to as \emph{strength}. 

Figure~\ref{fig:testsuite1} shows an example of a test suite for $t = 2$. 
The \emph{System Under Test (SUT)} assumed here is a printer.  
The SUT model has three parameters: 
Paper size, Feed tray, and Paper type. 
The set of possible values for these parameters are:  
\{B4, A4, B5\}, \{Bypass, Tray 1, Tray 2\}, and \{Thick, Normal, Thin\}. 
As can be seen from the figure, every pair-wise combination 
occurs in at least one of the test cases in the test suite. 
Note that the test suite is much smaller than the exhaustive test suite,  
which contains $3\times 3\times 3 = 27$ test cases. 

\begin{figure}[b]
	{\small
		\begin{tabular}{lll}
			\hline
			Paper size & Feed tray & Paper type \\ \hline \hline
			B4 & Bypass & Thick  \\
			B4 & Tray 1 & Thin   \\
			B4 & Tray 2 & Normal \\
			A4 & Bypass & Normal \\
			A4 & Tray 1 & Thick \\
			A4 & Tray 2 & Thin  \\
			B5 & Bypass & Thin \\
			B5 & Tray 1 & Normal \\
			B5 & Tray 2 & Thick \\  \hline
		\end{tabular}
	}
	\caption{Test suite of strength $t = 2$. No constraints are considered.} 
	\label{fig:testsuite1}
\end{figure}

This SUT example implicitly assumes that every possible test case can be executed. 
Real-world systems, however, often have \emph{constraints} concerning executable 
test cases. 
In such a case, all test cases must satisfy the given constraints
to execute.  
In addition, some interactions may become inherently not testable. 
Constraint handling is necessary to solve these problems in the process of 
test case generation. 
Though this might look an easy task, 
the efficiency of constraint handling sometimes 
greatly affects the whole performance of test case generation. 

It is easy to decide whether a test case satisfies the constraints, 
because it suffices to check if the test case satisfies each constraint one by one. 
The difficulty of constraint handling stems from the need 
to determine whether an interaction can be testable or not, 
or equivalently whether an incomplete test case with unspecified values 
can be extended to a full-fledged test case that satisfies the constraints. 
For example, consider the following constraints.
\begin{itemize}
	\item
	If the Paper size is B4, then the Feed tray must be Bypass. 
	
	\item
	If the Feed tray is Bypass, then the Paper type must not be Thick. 
	
\end{itemize}
In this case, B4 and Thick cannot occur in the same test case and thus 
their interaction is not testable, 
because B4 requires Bypass which in turn prohibits Thick. 
However, this cannot be decided by analyzing each of the two constraints in isolation. 
We say that an interaction (partial test case) or test case is \emph{invalid} if 
it cannot be tested due to the constraints. 

When constructing a test suite for CIT, 
such validity checks need to be performed a large number of times. 
As a result, the time used for validity checks is often dominant in 
the whole running time required to generate a test suite. 
Although modern test case generation tools are usually able to handle 
real-world problems in practical running times, problems sometimes 
arise that take very long time to solve because of complex constraints. 
The existence of such hard problems can severely hinder the practical usefulness of CIT. 

In fact, we started our study on constraint handling in response to the 
demand from our industry collaborators who were not satisfied with 
a test case generation tool that was then popular. 
Their complaint was that the execution time of the tool was 
sometimes too long to be useful in practice, especially the SUT involves many constraints. 
To achieve better performance, we adopted \emph{Binary Decision Diagrams 
	(BDDs)}~\citep{Akers:1978:BDD:1310167.1310815,Bryant:1986:GAB:6432.6433} 
as a key data structure on which constraint handling is 
performed. 
A BDD is a data structure that can compactly represent a Boolean function.  
We developed a test generation tool named CIT-BACH using  BDDs.\footnote{http://www-ise4.ist.osaka-u.ac.jp/\~{}t-tutiya/CIT}
Our attempt was successful in the sense that our tool has been used in industry for years. 
However, the performance of BDD-based constraint handling 
has not been systematically evaluated so far. 
Because of the lack of studies, it has been not clear whether 
BDDs are indeed useful or other approaches perform better. 

Aimed at filling this lack, this paper first 1) describes the 
BDD-based constraint handling approaches and then 2) compares the 
performance between these BDD-based approaches and other approaches. We 
consider two different BDD-based approaches. The first approach is a 
natural and straightforward one.  It performs the check by 
computing the conjunction of Boolean functions representing the 
constraints and the given partial test case. The second approach uses 
the algorithm that we developed for constraint handling in 
the CIT-BACH tool.  In this approach, a BDD is constructed that 
represents all valid full and partial test cases. Once such a BDD has 
been created, the check can be performed simply by traversing the BDD 
from the root node to a terminal node, thus avoiding any further BDD 
operations. Although the basic form of this approach has been used in 
our tool for some years, this paper describes it for the first time, 
together with some optimization techniques. 

To compare these approaches with other constraint handling approaches, 
we developed a new program that combines the IPOG algorithm~\citep{DBLP:conf/ecbs/LeiKKOL07,DBLP:journals/stvr/LeiKKOL08} and the BDD-based constraint handling approaches. 
The IPOG algorithm is a well-known test suite generation 
algorithm and adopted by ACTS~\citep{DBLP:conf/icst/YuLKK13}, one of the state-of-the-art tools. 
Since our program and ACTS shares the same algorithm 
except in constraint handling, using the two programs allows us to compare 
the BDD-based constraint handling approaches and other approaches 
employed by ACTS. 
We conducted an experiment campaign where 
we applied the two tools to a collection of various problem instances.   
This collection includes 62 problem instances, 
all taken from previous studies in the field of CIT 
except one that was provided by our industry collaborators.

The rest of the paper is organized as follows. 
Section~\ref{sec:model} describes the model of SUTs. 
Section~\ref{sec:handling} describes the outline of test case generation algorithms 
and how constraint handling is needed in the process of test case generation. 
Section~\ref{sec:bdd} shows how to represent the set of valid test cases using a BDD. 
Section~\ref{sec:details} describes the two BDD-based constraint handling approaches 
which use the BDD representation of valid test cases described in Section~\ref{sec:bdd}. 
Section~\ref{sec:evaluation} presents the results of experiments. 
Section~\ref{sec:relatedwork} presents a concise survey of related work.
Section~\ref{sec:threats} describes the potential threats to validity. 
Finally we conclude this paper in Section~\ref{sec:conclusion}.

\section{Model}\label{sec:model} 

We assume that the SUT has $n$ parameter variables 
(or parameters for short), $P_1, ..., P_n$. 
Parameter variable $P_i$ has its finite domain $D_i$ from which a value for 
the parameter is drawn.  
For simplicity of presentation, we assume that 
$D_i = \{0, 1, ..., |D_i| -1\}$. 

A \emph{test case} is an $n$-tuple such that its $i$th element 
is a value for $P_i$. 
Hence the set of test cases is $D_1 \times D_2 \times ... \times D_n$. 

In practice, a test case may need to satisfy some condition 
to be executed. 
\emph{Constraints} over parameter values mean such a condition. 
The constrains are formally represented as 
a Boolean-valued function $F$ over the parameter variables:  
\[
F: D_1 \times D_2 \times ... \times D_n \rightarrow \{0, 1\}
\]
where a test case $t$ satisfies the constraints if 
and only if $F(t) = 1$. 
We say that a test case is \emph{valid} if and only if 
it satisfies the constraints. 
We assume that $F$ is given as a formula composed of 
the Boolean connectives (such as $\neg$, $\land$, $\lor$ and $\Rightarrow$) 
and terms in the forms of $P_i = v \ (v \in D_i)$ and $P_i = P_j$. 

For instance, the SUT model shown in Section~\ref{sec:intro} is expressed using 
$n=3$ parameter variables $P_1, P_2, P_3$ which respectively represent 
Paper size, Feed tray and Paper type. 
The domains of the parameter variables are $D_1, D_2$ and $D_3$ where 
$D_1 = D_2 = D_3 = \{0, 1, 2\}$. 
We assume that each integer in the domains corresponds to a parameter value 
as follows:

\begin{quote}
	\begin{tabular}{llll}
		$D_1$ & 0: B4     & 1: A4     & 2: B5 \\
		$D_2$ & 0: Bypass & 1: Tray~1 & 2: Tray~2 \\
		$D_3$ & 0: Thick  & 1: Normal & 2: Thin\\
	\end{tabular}
\end{quote}

\noindent 
The Boolean-valued function representing the constraints of the example is as follows:
\[
F := (P_1 = 0 \Rightarrow P_2 = 0) \land (P_2 = 0 \Rightarrow \neg (P_3 = 0))
\]
A test case is a vector of length~3 whose element is either 0, 1 or~2. 
For example, (1, 0, 2) or (2, 0, 0) are test cases. 
The former test case is valid, whereas the latter is not valid.

\begin{figure}
	{\small
		\begin{tabular}{lll}
			\hline
			Paper size & Feed tray & Paper type \\ \hline \hline
			Bypass & B4 & Thin \\
			Bypass & A4 & Normal \\
			Tray1 & A4 & Thick \\
			Tray2 & A4 & Thin \\
			Bypass & B5 & Normal \\
			Tray1 & B5 & Thin \\
			Tray2 & B5 & Thick \\
			$-$ & B4 & Normal \\
			Tray1 & $-$ & Normal \\
			Tray2 & $-$ & Normal \\ \hline 
		\end{tabular}
	}
	\caption{Test suite constructed by IPOG for the running example (strength $t=2$). 
		Although there are three partial test cases, all pairwise interactions have already 
		been covered. 
		The process of assigning values to the entries with a $-$ is omitted in Figure~\ref{fig:ipogc} 
		for simplicity.
	}
	\label{fig:testsuite2}
\end{figure}

\begin{figure*}[t]
	{\small 
		\begin{algorithmic}[1]
			\STATE Sort the parameters in non-increasing order of $|D_i|$ \\
			\quad \COMMENT{For simplicity we assume $P_1, ..., P_n$ are sorted in this order.}
			\STATE $TestSuite \leftarrow \emptyset$
			\STATE \COMMENT{Cover all value combinations of the first $t$ parameters} \label{ln:firstts}
			\FOR{each combination $I$ of values of the first $t$ parameters}
			\IF{\texttt{isValid(}$I$\texttt{)}}
			\STATE $TestSuite \leftarrow TestSuite \cup \{T\}$
			\ENDIF
			\ENDFOR \label{ln:firstte}
			\STATE \COMMENT{Add remaining $n-t$ parameters} 
			\FOR{each $i = t+1, ..., n$}
			\STATE \COMMENT{Enumerate valid $t$-way value combinations of $P_i$ and $P_{i_1},...,P_{i_{t-1}}$}
			\STATE{$\pi \leftarrow \emptyset$}
			\FOR{each $t$-way combination $I$ of values of $P_i$ and $P_{i_1},...,P_{i_{t-1}}$ \label{ln:vchecks}
				such that $1 \leq i_1 < ... < i_{t-1} \leq i -1$} 
			\IF{\texttt{isValid(}$I$\texttt{)}}
			\STATE $\pi \leftarrow \pi \cup \{I\}$
			\ENDIF
			\ENDFOR \label{ln:vchecke}
			\STATE \COMMENT{Horizontal growth}\label{ln:horizs}
			\FOR{each $T = (v_1,..,v_{i-1})$ in $TestSuite$} 
			\STATE Choose a value $v_i$ of $P_i$ and replace $T$ with $T' = (v_1,...,v_{i-1},v_i)$
			so that \texttt{isValid(}$T'$\texttt{)}=true and $T'$ covers the most number of combinations of values in $\pi$  \label{ln:vcheck1} \\
			\STATE Remove the combinations of values covered by $T'$ from $\pi$ 
			\ENDFOR \label{ln:horize}
			\STATE \COMMENT{Vertical growth} \label{ln:verts}
			\FOR{each combination $I$ in $\pi$}
			\IF{there exists $T \in TestSuite$ that can be changed to $T'$ such that it covers 
				both $T$ and $I$ and \texttt{isValid(}$T'$\texttt{)}=true} \label{ln:vcheck2}
			\STATE Replace $T \in TestSuite$ with $T'$
			\ELSE
			\STATE $TestSuite \leftarrow TestSuite \cup \{I\}$
			\ENDIF
			\ENDFOR \label{ln:verte}
			\ENDFOR
		\end{algorithmic}
	}
	\caption{IPOG-C algorithm (based on \citep{ACTS2013})}
	\label{fig:ipogc}
\end{figure*}

In addition to full test cases, 
we often need to consider ``incomplete'' test cases where 
there are some parameters whose value is yet to be fixed. 
We call such a vector a \emph{partial test case}. 
More precisely, a partial test case is an $n$-tuple such that 
1) the $i$th element is either a value in $D_i$ or special symbol 
$-$ representing that the value is unspecified, 
and 2) at least one parameter has a $-$. 
A partial test case $t$ is \emph{valid} 
if and only if $t$ can be extended to 
a valid test case by assigning values to all the positions 
on which $-$ is placed. 
For example, $(1, 1, -)$ and $(0, -, 0)$ are partial test cases for the 
running example. 
The former partial test case is valid. 
On the other hand, the latter one is not valid (invalid), 
as explained in Section~\ref{sec:intro}. 

\section{IPOG test case generation algorithm}\label{sec:handling}

Algorithms for constructing test suites for CIT have been widely studied. 
Roughly they can be classified into two groups: mathematical and 
computational. Mathematical algorithms use properties of mathematical 
elements, such as finite fields, to generate test suites. Basically it 
is difficult for these algorithms to deal with SUTs with constraints. As 
a result, there are only few test generation tools that use this group 
of algorithms. 

Test case generation tools that have industrial strength usually use 
computational algorithms. In this paper we focus on IPOG~\citep{DBLP:conf/ecbs/LeiKKOL07,DBLP:journals/stvr/LeiKKOL08}, because 
ACTS~\citep{DBLP:conf/icst/YuLKK13}, 
which is a state-of-the art test generation tool for CIT, uses this 
algorithm. 
As described in Section~\ref{sec:evaluation}, we develop a program that employs IPOG 
and constraint handling techniques using BDDs. 
This allows us to compare BDD-based techniques with different 
constraint handling techniques that are adopted by ACTS. 


To be precise, the IPOG algorithm is not a single algorithm; rather it encompasses a few variants.  
The IPOG algorithm with the constraint handling capability is called IPOG-C~\citep{ACTS2013}; however, as our concern here is about constraint handling, we call IPOG-C simply IPOG. 
Figure~\ref{fig:ipogc} shows the pseudo-code of this algorithm. 
The algorithm constructs a test suite as follows. It starts with 
a set of $t$ parameters and constructs a test suite for that parameter set (lines~\ref{ln:firstts}--\ref{ln:firstte}). 
Then, it adds a new parameter and decides a value on that parameter for 
every existing test case (horizontal growth, lines~\ref{ln:horizs}--\ref{ln:horize}). If some $t$-way 
combinations are still missing in the test suite, the algorithm adds new 
test cases to the test suite to cover all such combinations (vertical 
growth, lines~\ref{ln:verts}--\ref{ln:verte}). As a result of iterating horizontal growth and vertical growth 
for all remaining parameters, a test suite for all parameters will be eventually 
constructed. Figure~\ref{fig:testsuite2} show the test suite constructed 
by our IPOG implementation for the running example. 

In this pseudo-code, we specify the points where constraint handling is called for by 
function \texttt{isValid()}. 
This function takes a full or partial test case $t$ as an argument and returns true if $t$ 
is valid; it returns false, otherwise.  
Note that in Figure~\ref{fig:ipogc}, a value combination of parameters is treated 
as a partial test case that covers the combination and has a $-$ on all remaining parameters.  


As can be seen in Figure~\ref{fig:ipogc}, the validity check is performed 
a number of times in the process of constructing a single test suite:  
All $t$-way combinations of parameter values need to undergo  
the validity check (lines~\ref{ln:firstts}--\ref{ln:firstte}, \ref{ln:vchecks}--\ref{ln:vchecke}). 
In addition, when deciding a value on parameter $P_i (i> t)$ of a test case, the validity 
check is performed (lines~\ref{ln:vcheck1} and~\ref{ln:vcheck2}). 
Of course, some optimization can be made. 
For example, no validity check is required for any interaction among a subset of parameters
if none of these parameters occur in the constraints.  
Although such optimizations are possible, the efficiency of constraint handling 
still has substantial effects on the performance of test case generation.  
In Section~\ref{sec:evaluation} we show this through experiments.

\section{BDD representations of valid test cases}\label{sec:bdd} 

In this section, we describe the concept of a BDD and how it can be used to 
represent the set of valid test cases. 
The BDD representation of valid test cases provides the basis for  
the validity checking approaches described in Section~\ref{sec:details}. 

\subsection{Binary Decision Diagrams} \label{subsec:bdd}

A BDD is a data structure that represents a Boolean function~\citep{Bryant:1986:GAB:6432.6433}. 
It is a rooted directed acyclic graph with two types of nodes, 
terminal nodes and nonterminal nodes. Each nonterminal node is 
associated with a Boolean variable and has two successor nodes. Each 
edge is labeled by either 0 or 1. A terminal node is labeled by F (false) 
or T (true). 
A BDD is \emph{ordered} if the Boolean variables occur in the same order  
along any path from the root to one of the terminal nodes. 
An ordered BDD can be transformed into a \emph{reduced} ordered BDD (ROBDD)
by repeatedly applying reduction rules until no further applications are possible. 
The application of these reduction rules removes all redundant isomorphic subgraphs and non-terminal nodes 
whose successor nodes are identical. 
For example, there are no more than two terminal nodes in an ROBDD. 
When the ordering of Boolean variables is fixed, an ROBDD is unique for a given Boolean function. 
Although there are several variants of BDDs, ROBDDs are now almost a synonym 
of BDDs; thus we call an ROBDD simply a BDD. 

Figures~\ref{fig:bdd1} and~\ref{fig:bdd2} schematically show BDDs for 
Boolean functions with six variables $x_1, x_2, ..., x_6$. 
Edges labeled by 0 are represented as dotted lines, 
whereas those labeled by 1 are represented as solid lines. 

Given a valuation of the Boolean variables, 
that is, an assignment of truth values to the Boolean variables, 
the value of the Boolean function for the valuation is determined by 
traversing the BDD from the root according to the valuation. 
Specifically, when the current visiting node is associated with 
a Boolean variable $x_i$ and $x_i$ takes 0 (or 1) in the valuation, 
then the traversal proceeds by taking the edge labeled by 
0 (or 1) to one of its successor nodes.  
If the terminal node reached is labeled by F (or T), then 
the value of the function is 0 (or 1). 
This check takes time linear in the length of the traversed path 
which is at most equal to the number of the Boolean variables. 


From two BDDs that represent Boolean functions $b$ and $b'$, 
another BDD that represents Boolean function $b \langle op \rangle  b'$ 
can be obtained with the algorithm $Apply$~\citep{Bryant:1986:GAB:6432.6433}, 
where  $\langle op \rangle$ can be 
any binary Boolean connective, such as $\land$ and $\lor$. 
The time complexity of $Apply$ is at most linear in the product of 
the numbers of nodes in the BDD representations of $b$ and $b'$.  
The negation of a Boolean formula can be obtained even easier: 
it suffices to swap the two terminal nodes. 

\subsection{Constructing a BDD that represents the set of valid test cases}\label{subsec:ordinary}

\begin{figure}
	\includegraphics[scale=0.6]{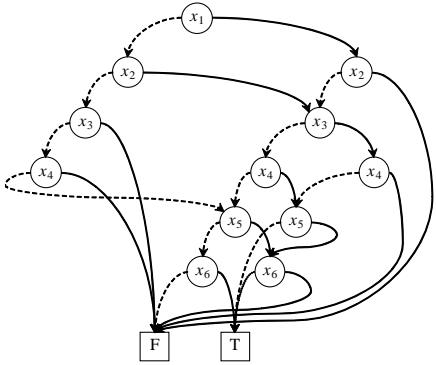}
	\caption{BDD representing the set of valid test cases. 
		The dotted lines and solid lines represent edges labeled with 0 and 
		edges labeled with 1.} 
	\label{fig:bdd1}
\end{figure}


Figure~\ref{fig:bdd1} shows a BDD representing the set of valid test cases
for the running example. 
Here the value of each parameter $P_i$ is represented by 
two Boolean variables $x_{2i-1}$ and  $x_{2i}$ with $x_{2i-1}$ being 
the least significant bit.
For example, consider test cases $(1, 0, 2)$ 
and $(2, 0, 0)$, which represent (A4, Bypass, Thin) and (B5, Bypass, Thick), 
respectively. 
These test cases are represented by $x_1, ..., x_6$ as follows.
\[
(1,0,2): \underbrace{1 0}_{x_1 x_2} \underbrace{0 0}_{x_3 x_4}  \underbrace{0 1}_{x_5 x_6}  
\quad 
(2,0,0): \underbrace{0 1}_{x_1 x_2} \underbrace{0 0}_{x_3 x_4}  \underbrace{0 0}_{x_5 x_6}  
\]
Here we write the least significant bit at the leftmost end for each parameter 
so that the above ordering of the Boolean variables can coincide with that of the BDD. 
Whether a test case is valid or not can be easily determined 
by traversing the BDD from the root node according to the 0-1 values of these Boolean variables.  


It should be noted that this BDD provides the Boolean function 
representation of $F$, i.e., the function that determines valid test 
cases (see Section~\ref{sec:model}). 
We let $f$ denote this Boolean function.  
This function is of the form:
$f: \{0, 1\}^{M} \rightarrow \{0, 1\}$,  
where $M$ is the number of Boolean variables. 
This function is obtained by computing the logical AND of 
the Boolean functions representing the condition that 
the value for $P_i$ is less than $|D_i|$ and 
the Boolean function representation of 
the SUT's constraints among the parameter values. 

The Boolean function representing $P_i < |D_i|$
is obtained from the binary representation of $|D_i|-1$. 
For example, suppose that $D_i = \{0, 1, 2\}$. 
Then $D_i$ is represented by two Boolean variables:  
$x_{i1}$ and $x_{i2}$. 
From the binary representation of 2, i.e., 10, 
one can derive the following function to ensure that 
the value for $P_i$ is at most 2.  
\[
x_{i2} \Rightarrow \neg x_{i1}
\]

The given SUT's constraints are straightforwardly represented 
as a Boolean function, because the Boolean function 
can be obtained by simply converting the constraints 
into binary representation.  
For example, consider the running example. 
Constraint $P_1 = 0 \Rightarrow P_2 = 0$ is represented as 
\[
\neg x_1 \neg x_2 \Rightarrow \neg x_3 \neg x_4 
\] 
Similarly, $P_2 = 0 \Rightarrow \neg (P_3 = 0)$ is represented as: 
\[
\neg x_3 \neg x_4 \Rightarrow \neg (\neg x_5 \neg x_6)
\]
As a result, the set of valid test cases defined by $F$ is represented 
as the following Boolean function $f$. 
\[
\begin{array}{ll}
& (x_2 \Rightarrow \neg x_1) \land (x_4 \Rightarrow \neg x_3) \land 
(x_6 \Rightarrow \neg x_5)  \\ 
\land & (\neg x_1 \neg x_2 \Rightarrow \neg x_3 \neg x_4) \\
\land & 
(\neg x_3 \neg x_4 \Rightarrow \neg (\neg x_5 \neg x_6))
\end{array}
\]
The BDD shown in Figure~\ref{fig:bdd1} exactly represents this Boolean function. 
All these operations over Boolean functions can be performed by manipulating their BDD representations. 

\section{Checking the validity of partial test cases}\label{sec:details}

As stated in the previous section, once the BDD representing 
the set of all valid test cases has been constructed, 
whether a given test case is valid or not can be easily checked 
by traversing the BDD from the root according to the binary 
representation of the test case. 
On the other hand, the check whether a partial test case is valid or not 
cannot be performed only by traversing the BDD. 

In this section, we present two approaches to validity checking for 
partial test cases. 
The first approach is natural and straightforward. 
It computes the conjunction, 
that is, logical AND of Boolean functions representing the constraints and the given 
partial test case and then checks if the Boolean function represented 
by the resulting BDD is a constant false, in which case the partial 
test case is invalid. 

The second one is a novel approach. 
In the approach, the BDD representing the space of valid test cases 
is modified so that it can represent both valid test cases and 
valid partial test cases. 
As a result, the validity of partial test cases can be checked 
by traversing the BDD from the root to the terminal nodes, just in 
the same way as the validity of full test cases are determined. 

\subsection{Approach~1: ANDing Boolean functions} 

The first approach requires building another BDD that represents a given partial test case 
and performing the AND operation over the BDD and the BDD representing $f$ 
(i.e., the BDD representing the set of valid test cases). 
If the resulting BDD represents a contradiction (constant false), 
then the partial test case is invalid; otherwise, it is valid. 
The number of Boolean variables used in this approach is 
$\sum_{i=1}^n \lceil \log_2 |D_i| \rceil$. 

The Boolean function that represents a partial test case 
is constructed by ANDing the binary representations of the parameter 
values that have already been fixed. 
For example, consider two partial test cases as follows: 
\[
(1,1,-): \underbrace{1 0}_{x_1 x_2} \underbrace{1 0}_{x_3 x_4}  \underbrace{- -}_{x_5 x_6}  
\quad 
(0,-,0): \underbrace{0 0}_{x_1 x_2} \underbrace{- -}_{x_3 x_4}  \underbrace{0 0}_{x_5 x_6}  
\]
Note again that we write the least significant bit at the leftmost end for each parameter.  
The Boolean functions that represent these partial test cases 
are: 
\[
(1,1,-): x_1 \neg x_2 x_3 \neg x_4 
\quad 
(0,-1,0): \neg x_1 \neg x_2 \neg x_5 \neg x_6 
\]

Let $f_T$ be a Boolean function representing a partial test case $T$. 
Since the Boolean function $f_T \land f$ represents the set of valid test cases 
that cover $T$, the validity check of $T$ can be implemented as follows: 
\[
\mathtt{isValid(}T\mathtt{)} = \left\{
\begin{array}{cl}
false  & f_T \land f = \mathrm{constant\ false} \\
true & \mathrm{otherwise}
\end{array}
\right.
\]
The BDD representing a constant false consists only of a single terminal 
node which is labeled with F; thus the time complexity of checking if the resulting BDD represents a constant 
false or not is $O(1)$. 
The worst-case time complexity of computing $f_T \land f$ is linear in 
the product of the sizes of the two BDDs.  
However, in practice the computation time is usually small, 
because $f_T$ has a very simple BDD representation, where 
there are only as many non-terminal nodes as Boolean variables occurring in $f_T$.  
In Section~\ref{sec:evaluation}, we experimentally evaluate the performance of this approach. 

\subsection{Approach~2: Constructing a BDD representing partial test cases}

\subsubsection{Outline}

\begin{figure}
	\includegraphics[scale=0.6]{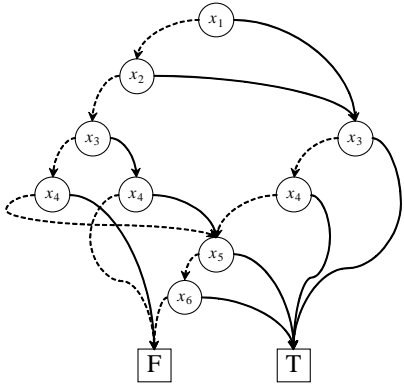}
	\caption{BDD representing all valid full and partial test cases}
	\label{fig:bdd2}
\end{figure}

The fundamental idea of this approach is to construct 
a BDD that represents the set of all valid full/partial test cases.
This allows us to determine very quickly whether a given partial test case 
is valid or not. 

Now consider extending the domain $D_i$ of parameter $P_i$ by 
adding the element $-$ which represents that the value 
of the parameter is unspecified. 
Then the set of valid full and partial test cases is represented 
in the form of a Boolean-valued function as follows: 
\[
G: D_1\cup \{-\} \times D_2\cup \{-\}
\times ... \times D_n\cup \{-\} \rightarrow \{0, 1\}
\]
where a full or partial test case $T$ is valid if and only if $V(T) = 1$. 

To encode the domain of the function $V$ in bit strings, 
the special value $-$ is represented as $11...1$, whereas the values 
in $D_i$ are represented in the usually binary form, just as in the 
ordinary BDD representation.  

Figure~\ref{fig:bdd2} shows the BDD that represents $G$ for the 
running example. 
Partial test cases are represented using valuations to Boolean variables $x_1, ..., x_6$, 
just as full test cases. 
Below are examples. 
\[
(1,1,-): \underbrace{1 0}_{x_1 x_2} \underbrace{1 0}_{x_3 x_4}  \underbrace{1 1}_{x_5 x_6}  
\quad 
(0,-,0): \underbrace{0 0}_{x_1 x_2} \underbrace{1 1}_{x_3 x_4}  \underbrace{0 0}_{x_5 x_6}  
\]

Given a binary representation of a test case or partial test case, checking if 
it is valid or not can be done simply by traversing the BDD from the root node to 
one of the terminal nodes. 
The time required for the validity check based on the BDD is linear in 
the length of the traversed path. 
As the path length is at most the number of Boolean variables, 
the time required for the validity check is linear in the number of 
parameters and logarithmic in the number of values of possible parameter values.


\subsubsection{Constructing a BDD representing valid full and partial test cases}

\begin{figure*}
	\begin{algorithmic}[1] 
		\STATE $g$ $\leftarrow$ $f$: Boolean function that represents $F$  
		\quad \COMMENT{Initially $g$ represents the set of all valid full test cases.} 
		\FOR{each parameter variable $P_i$} 
		\STATE $cube$ $\leftarrow$ $x_{M_{i-1}+1} x_{M_{i-1}+2} ... x_{M_i}$ 
		\quad \COMMENT{$x_{M_{i-1}+1}, ... ,x_{M_i}$ are the Boolean variables representing $P_i$.} \label{ln:cube}
		\STATE $tmp$ $\leftarrow$ $Exists(cube, g)$ 
		\quad \COMMENT{$tmp = \exists x_{M_{i-1}+1}...\exists x_{M_i} g$} 
		\STATE $h$ $\leftarrow$ $Apply(\land, tmp, cube)$;  
		\quad \COMMENT{$h = \exists x_{M_{i-1}+1}...\exists x_{M_i} g \ \land \ x_{M_{i-1}+1} x_{M_{i-1}+2} ... x_{M_i}$} \label{ln:and}
		\STATE $g$ $\leftarrow$ $Apply(\lor, f, h)$ 
		\quad \COMMENT{$g$ is replaced with $g \lor h$}  \label{ln:or}
		\ENDFOR
	\end{algorithmic}
	\caption{BDD modification algorithm. The algorithm converts the BDD 
		representing $F$ into another BDD representing $G$, where $F$ and $G$ respectively express 
		the set of valid full test cases and the set of valid full and partial test cases.}
	\label{fig:modifyBdd}
\end{figure*}

Below we show how to construct the BDD that represents 
the set of all valid full and partial test cases. 
This process consists of two steps. 
In the first step, the BDD that represents $F$, or equivalently  
the set of all valid full test cases is constructed. 
This is the same as in Approach~1, except that the number 
of Boolean variables required for parameter $P_i$ is now 
$\log_2(|D_i| + 1)$, 
instead of $\log_2 |D_i|$, because we need to distinguish the special value 
$-$ from the values in $D_i$.  
Let Boolean variables $x_{M_{i-1}+1}, x_{M_{i-1}+2}, ...,  x_{M_i}$ 
be these variables where $M_i$ is defined as follows:
\[
M_i = 
\left\{
\begin{array}{ll}
0 & i = 0 \\
M_{i-1} + \lceil \log_2 (|D_i| + 1) \rceil & 1 \leq i \leq n 
\end{array}
\right.
\]
Hence the total number of Boolean variables equals $M_n$. 
For the running example, 
$M_0 = 0, M_1 = 2, M_2 = 4$ and $M_3 = 6$, 
as $P_1, P_2$ and $P_3$ are represented using 
$\{x_1, x_2\}$, $\{x_3, x_4\}$ and $\{x_5, x_6\}$, respectively. 

In the second step, the BDD is modified so that a new 
BDD that represents $G$, that is, the set of all valid full and partial test cases 
is obtained. 
This modification is implemented by iteratively transforming $f$ such that a Boolean 
function that represents $G$ can be eventually obtained. 
Let us define $G_0 := F$ and $g_0 := f$.  
The $i$th iteration computes $g_i$ from $g_{i-1}$. 
Finally, the result of the $n$th iteration, $g_n$, 
becomes the Boolean function that represents $G$. 


In addition to the basic Boolean operators, such as $\land$ or $\lor$, 
this step uses the existential quantifier $\exists$. 
The existential quantification is defined as follows:
\[
\exists x_j b(x_1, ...,x_N) = b_0 \lor b_1
\]
where $b(x_1, ...,x_N)$ is a Boolean function and $b_c = b(x_1,$ $...,$ $x_{j-1}$, $c$, $x_{j+1}$,$...$,$x_N)$ ($c\in \{0, 1\}$). 

When $m$ existential quantifiers are applied simultaneously, 
the resulting function is a conjunction of a total of $2^m$ functions, 
each obtained by assigning 0 or 1 to the variables quantified. 

The Boolean function $g_1$ is a Boolean function representation 
of the following function $G_1$:  
\[
G_1: D_1 \cup \{-\} \times D_2 \times ... \times D_n \rightarrow \{0, 1\}
\]
where 
$G_1(T) = 1$ if and only if $T$ is a valid full or partial test case 
in $D_1 \cup \{-\} \times D_2 \times ... \times D_n$. 
Note that here $T$ can have a symbol $-$ only on the first parameter. 
In other words, $G_1$ characterizes the set of all full and partial 
test cases such that a symbol $-$ may occur only on the first parameter. 

The difference between $G_0 (= F)$ and $G_1$ is that $G_1$ represents, 
in addition to all valid test cases which are represented 
by $G_0$, all valid partial test cases that have a $-$ exactly on the first 
parameter. 
Therefore, $g_1$, the Boolean function representation of $G_1$, 
can be obtained by composing the logical OR 
of $g_0$ and the Boolean function, denoted as $h_1$, that represents 
all valid partial test cases with a $-$ on the first parameter.
The Boolean function $h_1$ is obtained as follows:
\[
h_1 := \exists x_1 \exists x_2 ... \exists x_{M_1} g_0 \ 
\land \ x_1 x_2 ... x_{M_1}
\]
Note that $\exists x_1 \exists x_2 ... \exists x_{M_1} g_0(\mathbf{c}) = 1$  
for $\mathbf{c} \in \{0, 1\}^{M_n}$ if and only if
at least one valuation $\mathbf{c'}$ exists such that $g_0(\mathbf{c'}) = 1$ 
and $\mathbf{c'}$ has the same truth values as $\mathbf{c}$ for $x_{M_1 +1}, ..., x_{M_n}$. 
This means that $\exists x_1, \exists x_2, ..., \exists x_{M_1} g_0(\mathbf{c}) = 1$ 
if and only if there is a valid test case $T = ( v_1, v_2, ..., v_n )$
such that every $v_i$ with $i \geq 2$ is the parameter value represented by $\mathbf{c}$. 
The last conjunct $x_1 x_2 ... x_{M_1}$ becomes 1 
if and only if the valuation has 1 for $x_1, ..., x_{M_1}$, 
meaning that a $-$ is placed on the first parameter. 
As a result of the logical OR operation over these two Boolean functions, 
$h_1$ characterizes all partial test cases that have a $-$ exactly on the first parameter. 
Finally, $g_1$ is obtained as follows: 
\[
g_1 := g_0 \lor h_1 
\]

The above process can be generalized as follows. 
Let $G_i$ for $1 \leq i \leq n$ be the function 
\[
G_i: D_1 \cup \{-\} \times ... \times D_i \cup \{-\} 
\times D_{i+1} \times ... \times D_n \rightarrow \{0, 1\}
\]
such that 
$G_i(T) = 1$ if and only if $T$ is a valid full or partial test case 
belonging to the domain of $G_i$. 
Thus $G_i$ characterizes the set of valid full and partial test cases 
such that a $-$ can occur only on $P_1, ..., P_i$. 
Let $g_i$ be the Boolean function that represents $G_i$ 
following our encoding. 

Now suppose that $g_{i-1}$  $(1 \leq i \leq n)$ has already been obtained. 
One can compute $g_{i}$ from $g_{i-1}$ the similar way as in the case 
of computing $g_1$ from $g_0$. 
First, we compute the Boolean function, $h_i$, that characterizes the set of 
partial test cases such that a $-$ must occur on parameter $P_i$ and may occur on 
$P_1, ..., P_{l-1}$. This function is obtained as follows: 
\[
\begin{array}{lll}
h_i & :=  & \exists x_{M_{i-1}+1} \exists x_{M_{i-1}+2} ... \exists x_{M_{i}} g_{i-1} \\ 
&     & \land \ x_{M_{i-1}+1} x_{M_{i-1}+2} ... x_{M_{i}}
\end{array}
\]
Hence $g_i$ can be obtained as follows.
\[
g_i := g_{i-1} \lor h_{i}
\]
By repeating this process $n$ times, Boolean function $g_n$ that 
represents the set of all valid full and partial test cases is obtained.

Figure~\ref{fig:modifyBdd} shows the algorithm for computing $g_n$ from $f$. 
As stated in Section~\ref{subsec:bdd}, binary Boolean operations 
are performed by the algorithm $Apply$. 
In Figure~\ref{fig:modifyBdd}, $Apply(\land, b, b')$ (line~\ref{ln:and}) and $Apply(\lor, b, b')$ 
(line~\ref{ln:or}) stand for the logical OR 
and AND operations over two Boolean functions $b$ and $b'$ using the $Apply$ algorithm. 
Multiple existential quantification can be done using 
an algorithm usually called the $Exists$ algorithm~\citep{knuth},   
where the variables to be quantified are given as a BDD representing their conjunction 
(denoted by $cube$ at line~\ref{ln:cube} in Figure~\ref{fig:modifyBdd}).

\subsection{Optimizations}

In our current implementation of the two approaches, we make some optimizations 
shown below. 

\subsubsection{Ignoring unconstrained parameters}
The first optimization is to ignore parameters that do not occur in the 
constraints. 
This is based on the fact that the value on such parameters never 
affects the validity of a full or partial test case. 
Ignoring these unconstrained parameters reduces 
BDD size and in turn computation time. 
For example, suppose that the SUV has four parameters $P_1, P_2, P_3, P_4$ 
and that no constraints involve $P_3$. 
In this case, the number of Boolean variables required  is 
$\sum_{i\in \{1,2,4\}}\lceil \log_2 |D_i|\rceil$ (Approach~1) or 
$\sum_{i\in \{1,2,4\}}\lceil \log_2 (|D_i| + 1) \rceil$ (Approach~2). 
A given full or partial test case is treated as if $P_3$ did not exist. 
For example, $(0, 1, 2, 3)$, $(0, -, 2, 3)$ and $(0, 1, -, 3)$ are 
truncated by removing the value on $P_3$, thus resulting in 
$(0, 1, 3)$, $(0, -, 3)$ and $(0, 1, 3)$, respectively.  
Note that in this case, the validity check of $(0, 1, -, 3)$ 
can be performed in the same way as that of full test cases; 
that is, it can be done simply by traversing the BDD from the root 
to one of the terminal nodes. 

\subsubsection{Variable ordering}\label{subsubsection:ordering}
The second optimization concerns \emph{variable ordering}. 
The size of BDDs considerably varies depending on the order of Boolean variables. 
Since the problem of finding the optimal order that minimizes a BDD is NP-hard~\citep{Bryant:1986:GAB:6432.6433},  
many heuristics for variable ordering have been proposed, especially in the field of 
computer-aided design of digital circuits. 
A general rule of thumb to produce a small BDD is to keep related variables close.  
Our program performs static variable ordering as follows:
1) the order of test parameters are determined using the parameters' mutual distance in  
given constraints, and 
2) the Boolean variables representing the same parameter are placed consecutively, 
just as presented in Section~\ref{sec:bdd}. 


We define the distance between two parameters on a \emph{parse tree}. 
Our program supports an input language similar to other modern CIT tools, such as ACTS and PICT~\citep{Czerwonka06}, 
where a constraint is specified by a mathematical expression over 
parameters with Boolean operators (e.g. $\neg$, $\land$, $\lor$, $\Rightarrow$) 
and arithmetic comparators (e.g. $=$, $\neq$, $>$).
We first build parse trees for the expressions representing the constraints.
If there are more than one constraint, a single virtual parse tree is constructed by 
connecting the root nodes of multiple parse trees, each representing a constraint, with a new 
root node. 
Then, the distance between every pair of constrained parameters is computed. 
We define the distance between two parameters as the minimum distance between two nodes 
that correspond to the parameter pair.
Parameter ordering is computed based on the distance as follows. 
As the first parameter, the parameter that has the least sum of distances to other parameters 
is selected. Then, remaining parameters are selected one by one in such a way that when 
selecting a parameter, the sum of distances 
from the parameter to those already selected is minimized. 
Boolean variables for parameters that are ordered first are placed first near the BDD root. 

\subsubsection{Quantification order} \label{subsubsection:quantification}
The third optimization applies only to Approach~2.
It concerns the order of quantifying Boolean variables. 
In the presentation of Approach~2, we assumed that 
parameters $P_1, P_2, ... $ are encoded using Boolean variables arranged in the 
same order and stated that 
the Boolean variables are quantified according to this order. 
But this quantification process works for any order of parameters 
and, from our experience, the order may sometimes affect performance.   
This stems from the property of the $Exists$ algorithm. 
The worst case running time of $Exists$ is $O(B^{2^m})$ where 
$B$ is the number of nodes of the BDD and $m$ is the 
number of variables quantified if the quantification occurs near the root of the BDD. 
On the other hand, the running time is $O(B)$ if the quantification occurs 
near the terminal nodes~\citep{knuth}. 
In the next section, we empirically compare two different orders: 
\textbf{(DOWN)} $P_1, P_2, ..., P_n$ in which case 
quantifications occur from the top to the bottom of the BDD 
and \textbf{(UP)} $P_n, P_{n-1}, ..., P_1$ in which case 
quantifications occur from the bottom to the top. 


\section{Evaluation}\label{sec:evaluation}

\subsection{Experiment settings} 

This section shows the results obtained in experiments. 
Our research question is:
\begin{quote}
	How much do the BDD-based constraint handling techniques 
	improve the test generation time compared to other techniques? 
\end{quote}
To answer the question, we developed a test case generation 
program that implements the BDD-based constraint handling approaches and 
compare it to ACTS (version 3.0) with respect to running time. 

The reasons for choosing ACTS for comparison are manyfold. 
First, ACTS is a state-of-the-art tool for combinatorial interaction 
testing. It has been used for many industrial projects. 

The second reason is that 
ACTS implements two different constraint handling approaches 
and allows the user to select which one is used. 
The first approach is to use a Constraint Satisfaction Problem (CSP) 
solver~\citep{ACTS2013}. A CSP solver can be naturally used to determine the 
validity of a full or partial test case, because it suffices to 
check the satisfiability of the SUT constraints over test parameters 
and a constraint that represents the given test case. 
ACTS uses the CHOCO CSP solver~\citep{choco}. 
The second approach is to use the notion of minimum forbidden tuples (MFTs)~\citep{DBLP:conf/icst/YuDLKK15}.  
An MFT is a minimal combination of values that never occur in valid test cases. 
The MFT-based constraint handler computes all MFTs at the beginning of test 
case generation as follows.   
First, hidden forbidden tuples are obtained by manipulating forbidden tuples induced by each constraint  
with basic tuple operations, such as truncation and concatenation. 
Then those that are not minimal are removed.  
MFTs are used during test case generation whenever 
validity check needs to be performed. 

Another reason is that ACTS is proven to be already fast especially for 
large-size problems. In~\citep{DBLP:conf/kbse/0002BAKC16}, for example, 
ACTS is compared by a new algorithm proposed by Yamada et al. The 
results show that the new algorithm exhibits better performance for many 
problems when strength $t$ (the size of interactions to be covered) 
is two but ACTS often runs much faster when $t = 3$,  
especially for problems that take long time to solve. 

Our research question concerns the performance of constraint handlers 
and not the performance of tools. 
To enable fair comparison in this sense, our 
program implements IPOG which is the test case generation algorithm 
implemented by ACTS. 
We wrote the program in Java, the same language as ACTS, to mitigate 
the performance difference caused by programming languages.   
We used an open source library called JDD for BDD operations.\footnote{https://bitbucket.org/vahidi/jdd/wiki/Home}
JDD is also written purely in Java.

Boolean satisfiability (SAT) solving is sometimes used for constraint handling 
in previous studies. 
In this approach, the validity of a full or partial test case is determined by 
checking the satisfiability of a Boolean formula that is a conjunction of
the formula that represents the constraints and the one that represents the test case. 
To compare the SAT-based approach with others, we incorporated it in our program 
using Sat4j  (vrelease 2.3.4)\citep{DBLP:journals/jsat/BerreP10}, a well-known SAT solver written in Java. 
Modern SAT solvers, including Sat4j, only accept Boolean formulas in Conjunctive Normal Form (CNF). 
We used another Java library called Sufferon (version 2.0) to transform Boolean expressions representing constraints into CNF.\footnote{https://github.com/kmsoileau/Saffron-2.0}


As input instances, we used 62 SUT models, including:
\begin{itemize}
	\item 35 models from \citep{Cohen:2008}. 
	This set of models has widely used in 
	studies in the CIT field (e.g.~\citep{CASA,TCA2015}).  
	\item 15 models from  \citep{Johansen:2012}. 
	These models are derived from feature models of real-world systems. 
	\item 10 purely synthetic models from \citep{ACTS2013}. 
	
	\item One model from \citep{4090256}. This model is derived from TCAS, a well-known avionics system.
	\item One model (denoted ``web'') provided by our industry collaborators. 
	This model was applied to testing of a web-based application used in mobile 
	communication industry. 
\end{itemize}
We ran ACTS and our program for these instances with strength $t = 2$ and 
$t=3$ and measured execution time and the size of the resulting test suites. 
When executing these programs, we set the initial and maximum heap sizes 
of the Java virtual machine to 1 GByte. 

For each problem instance, we executed 12 runs. 
The execution time was averaged over 10 of the 12 runs 
excluding the smallest and largest values. 
The experiment was conducted using a Ubutu 18.04.1 LTS PC 
with an Intel Core i7-6700 CPU at 3.40 GHz with 8.00 GB memory. 

\subsection{Results} 
\begin{figure*}
	\includegraphics[scale=0.5]{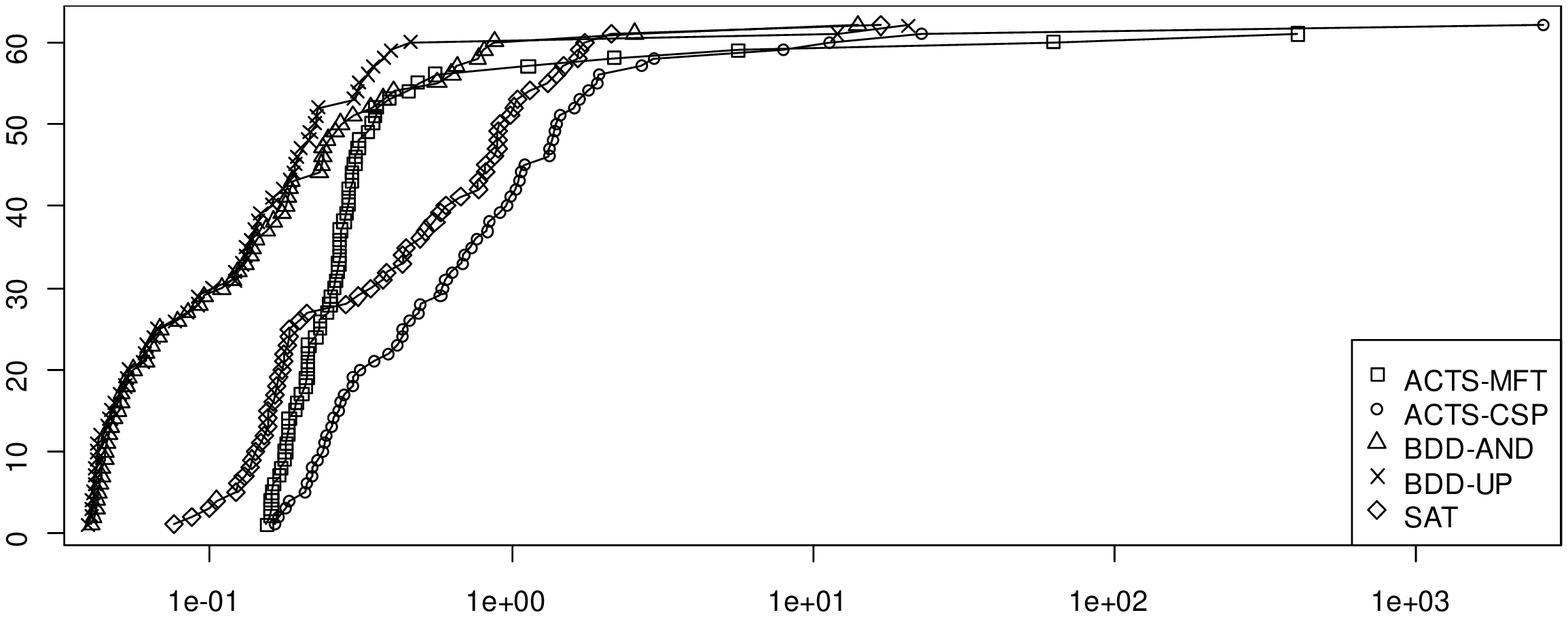}
	\includegraphics[scale=0.5]{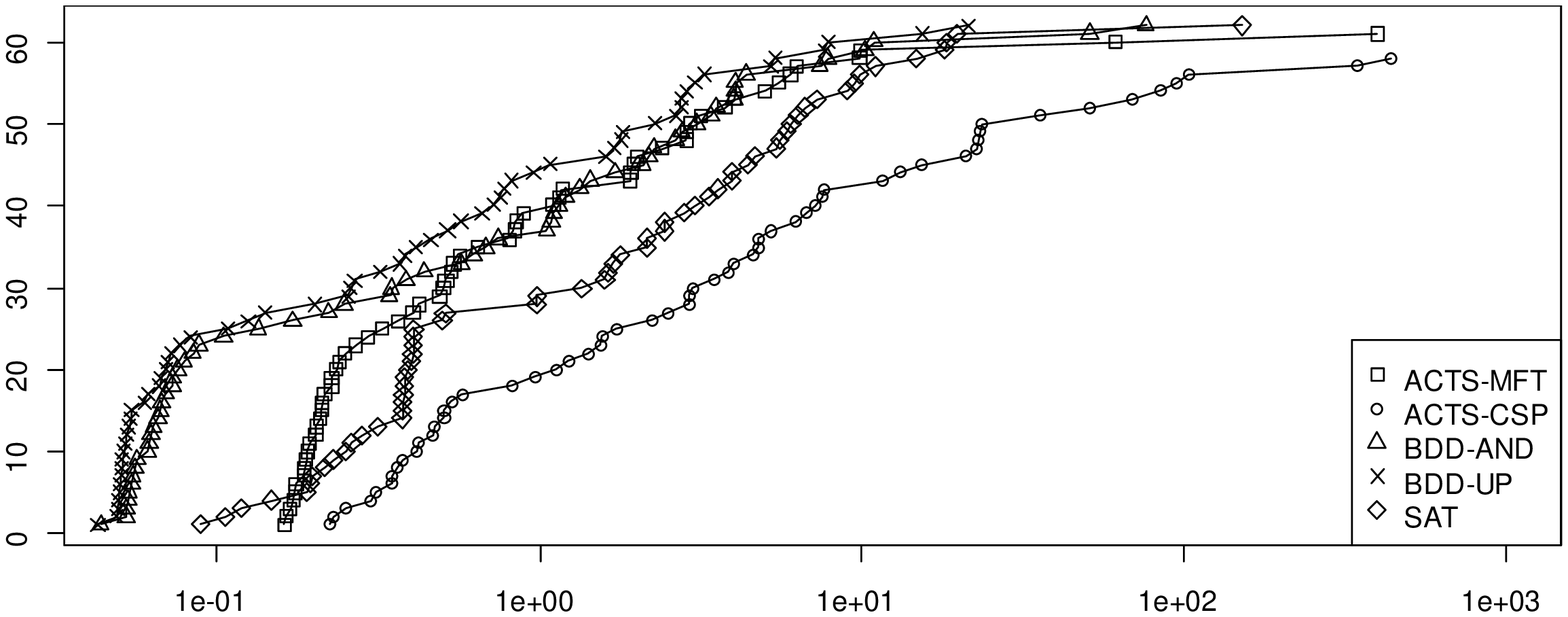}
	\caption{Cactus plots for five approaches (above: strength~$t = 2$, bottom: strength~$t = 3$). 
		Each curve represents, as a function of running time (in seconds), the number of problem instances solved 
		within the time. BDD-UP exhibited the best performance for both values of strength. 	
}
	\label{fig:cactus3}
	
\end{figure*}

\begin{table*}[htpb]
	\centering
	\caption{Running time (in seconds) for the hardest problems (strength $t = 3$).
		The shortest running time among the six approaches is underscored in bold format. }
	\label{tab:hardproblems}
	\begin{tabular}{l rrr rrr}
		\hline
		& ACTS- & ACTS- & BDD-  & BDD- & BDD- & \\
		& MFT & CSP & AND & DOWN & UP  & SAT     \\ \hline
		arcade\_game\_pl\_fm & 61.269   & NA       & 1.4349  & 0.8129   & \textbf{0.5684} & 2.161    \\
		benchmark5      & 6.0134   & 84.3539  & 7.828   & \textbf{5.3835}   & 5.4009  & 14.8036      \\
		benchmark10     & 3.7634   & 68.8019  & 4.3775  & 3.2494   & \textbf{3.2429}  & 11.0967      \\
		benchmark12     & 2.8616   & 23.0585  & 3.5192  & \textbf{2.6413}   & 2.6423  & 7.2384        \\
		benchmark18     & 2.9369   & 13.2259  & 4.0366  & \textbf{2.8233}   & 2.8669  & 9.016         \\
		benchmark19     & 9.8636   & 94.4125  & 10.2388 & 7.7881   & \textbf{7.7078}  & 18.3826       \\
		benchmark20     & 5.5407   & 342.7122 & 7.4293  & 5.307    & \textbf{5.2288}  & 18.2359      \\
		benchmark26     & 0.8882   & 11.6097  & 1.3248  & 0.7612   & \textbf{0.7546}  & 3.0411        \\
		benchmark28     & 9.9679   & 50.7824  & 10.9251 & 7.9583   & \textbf{7.8849}  & 19.9072       \\
		benchmarkapache & 4.055    & 4.6032   & 3.0835  & 2.7817   & \textbf{2.7571}  & 5.5723        \\
		benchmarkgcc    & 5.0245   & 6.255    & 4.0425  & \textbf{3.0095}   & 3.0324  & 5.8441        \\
		Berkeley        & 6.2977   & NA       & 2.1881  & 0.9352   & \textbf{0.8188}  & 3.5701        \\
		Gg4             & 398.7626 & 439.5144 & 0.3877  & 0.361    & \textbf{0.269}   & 0.9835      \\
		smart\_home\_fm & 0.2966   & 103.3322 & 0.2235  & 0.1418   & \textbf{0.1417}  & 0.3827         \\
		Violet          & \textbf{3.1556}   & NA       & 50.8893 & NA       & 21.5941 & 9.8687        \\
		web          & NA       & NA       & 75.7806 & 47.6126  & \textbf{15.4218} & 150.3782       \\ \hline
	\end{tabular}
\end{table*}

\begin{table}[htpb]
	\centering
	\caption{Running time (in seconds) for the hardest problems when constraints are ignored (strength $t = 3$).}
	\label{tab:noconstraints}
	\begin{tabular}{l rr} \hline
		& ACTS-NC & our IPOG \\ \hline
		arcade\_game\_pl\_fm &  0.1185  & 0.1969      \\
		benchmark5      &  3.1769  & 7.7908      \\
		benchmark10     & 1.703   & 4.518       \\
		benchmark12    & 1.557   & 4.0171      \\
		benchmark18    & 1.5735  & 3.6505      \\
		benchmark19    & 5.4875  & 11.1345     \\
		benchmark20      & 2.5186  & 6.4637      \\
		benchmark26      & 0.4076  & 0.874       \\
		benchmark28     &  5.1782  & 11.2293     \\
		benchmarkapache &  2.5305  & 4.3171      \\
		benchmarkgcc    &  4.3441  & 3.5499      \\
		Berkeley        &  0.1875  & 0.3071      \\
		Gg4             & 0.0497  & 0.0859      \\
		smart\_home\_fm & 0.044   & 0.0748      \\
		Violet          & 0.3798  & 0.5258     \\
		web          &  0.5944  & 2.4659      \\ \hline
	\end{tabular}
\end{table}

The results of this experiment are summarized in Figure~\ref{fig:cactus3} 
and Table~\ref{tab:hardproblems}.
(All raw data and the source code of our program are available on the WWW\footnote{https://osdn.net/users/t-tutiya/pf/IPOGwBDD/}.)
The approaches evaluated here are: 
ACTS with MFT-based constraint handling (denoted as ACTS-MFT),  
ACTS with CSP-based constraint handling (denoted as ACTS-CSP),  
and our program with BDD-based and SAT-based constraint handling (denoted as SAT). 
BDD-AND indicates Approach~1, whereas BDD-DOWN and BDD-UP mean 
Approach~2 with downward and upward variable quantification. 

Figure~\ref{fig:cactus3} presents the results with respect to running time 
in the form of cactus plots, which are often used to display 
performance of SAT solvers.  
The horizontal axes represent running time, whereas the vertical axes represent 
the number of problem instances (i.e. SUT models) solved within a given running time. 
Each curve is obtained for each approach by sorting the problem instances in ascending order of 
running time and by plotting a point for each problem instance on the horizontal 
coordinate corresponding to the running time.  
It should be noted that the order can be different for different approaches. 
As BDD-UP and BDD-DOWN showed similar performance for many of the instance, 
we omit the curves for the latter to avoid cluttering the plots. 

Of the five different approaches, BDD-UP showed the best performance. 
BDD-AND exhibited similar performance when strength $t$ is two, but 
the difference becomes clearer for the case $t=3$. 
This reason can be explained as follows. When $t = 2$, the number of times of validity checking is 
comparably small,   
because the number of pair-wise parameter value combinations is much smaller than that of 
three-way combinations. 
Nevertheless, Approach~2 (BDD-DOWN and BDD-UP) requires performing the costly BDD transformation 
regardless of the value of $t$, while Approach~1 (BDD-AND) does not undergo this step. 
The MFT-based approach comes next after BDD-AND, except that the SAT-based approach runs faster 
for easy problem instances. 
The CSP approach shows the worst performance among these approaches. 

These plots are convenient for comparing the performance of the different approaches; 
but it is not easy to comprehend performance difference for hard problem instances only from the plots.
Hence we sorted the problem instances in order of running time for the case $t=3$ and 
selected ten hardest ones for each approach. 
This ended up with a total of 16 instances. 
Table~\ref{tab:hardproblems} summarizes these instances and the running time for them. 

BDD-UP again exhibited best performance for these hard instances. 
BDD-DOWN showed similar performance to BDD-UP but significantly slowed down for some instances, 
specifically web and Violet.
As stated in Section~\ref{subsubsection:quantification}, we think that
this difference stems from the fact that the time complexity of the 
quantification algorithm varies depending on the position of the variables to be quantified.
The running time of BDD-AND was larger than that of the other BDD-based approach (Approach~2); 
but it showed better performance than the other non-BDD-based approaches.  
Overall, the BDD-based approaches exhibited better running time than the other approaches.

To show that the speedup was obtained from the efficiency of the 
constraint handling, we measured running time with constraint handling disabled. 
Table~\ref{tab:noconstraints} show the results of this experiment. 
The columns labeled with ACTS-NC and ``our IPOG'' show the running time 
of ACTS and our program when constraints are ignored. 
Although our program was slower 
than ACTS, both programs took a very short time to produce a test suite 
for most of the problems. Comparing the running time in the presence and 
absence of constraints, it can be observed that the time required for 
constraint handling is usually dominant in the overall running time when the SUT 
has constraints over parameters. 
This shows that the efficiency of constraint 
handling is of prime importance in improving the performance of test 
case generation. 
Exceptions are benchmark28, benchmarkapache, and benchmarkgcc. 
These problem instances have the common feature that  
ignoring constraints blows up the size of test suites and thus 
resulted in longer running time than when the constraints are taken into account.


Finally we touch on the size of resulting test suites. 
The IPOG implementation in both programs are deterministic 
(except for the process of filling $-$ entries, which was disabled in the experiment); thus 
each program always produces the same test suite for the same input.  
However, the test suites generated by the two algorithm for the same input are usually different.
This can be explained by that 
the level of abstraction of the IPOG algorithm leaves much room for  
different implementation variants. 
For example, when deciding a value on a parameter for an existing partial test case, 
IPOG selects the value that covers the most number of $t$-way parameter 
value combinations that are yet to be covered (line~\ref{ln:vcheck1} in Figure~\ref{fig:ipogc}).  
If there are more than one such value, then the tie must be broken somehow; 
but how to do so is not specified by the algorithm.
Nevertheless, the difference in test suite size was small: 
For 61 SUT models obtained from the 62 models by excluding the one that ACTS failed to 
handle, the relative difference in size (i.e. the number of test cases) was 0.043 for the case $t=2$
and 0.017 for the case $t=3$ on average. 
Thus, the difference in test suite size does not affect the above conclusions 
with respect to the performance of constraint handling.

\section{Related Work}\label{sec:relatedwork}

Constraint handling has been an issue in CIT since early studies in this field~\citep{Tatsumi87,264150}. 
There is even a systematic literature review on this particular topic~\citep{8102999}. 
Recent studies that address constraint handling often use 
solvers for combinatorial problems, such as Boolean satisfiability (SAT). 


Meta-heuristic algorithms, such as simulated annealing, are known to be 
effective for constructing small test suites for CIT, although they 
tend to consume relatively large amount of running time.
CASA~\citep{CASA} and TCA~\citep{7372037} are recent tools that adopt algorithms of this type.  
Both tools use a SAT solver for validity checking. 
In an early stage of our study, we experimented on CASA by modifying its SAT-based 
constraint handling module with a BDD-based one written in C++ to see the feasibility of the use of BDDs. 
We reported preliminary performance results in a Japanese article~\citep{sakano} but did not describe the details of 
the BDD-based approach itself. 


The one-test-at-a-time greedy algorithm is a class of algorithms which construct 
a test suite by repeatedly adding a test case to the current test suite 
until all necessary interactions are covered. 
PICT of Microsoft~\citep{Czerwonka06} and CIT-BACH are examples of tools 
that adopt algorithms of this class. 
Some recent studies on one-test-at-a-time greedy algorithms use solvers not only 
for validity checks but also for helping generate good test cases. 
Studies in this line of research include~\citep{Cohen:2008,6571642,DBLP:conf/kbse/0002BAKC16}. 
In contrast, neither PICT nor CIT-BACH uses combinatorial solvers. 
PICT uses a technique similar to the MFT-based one, while CIT-BACH uses the BDD-based 
approach as stated in Section~\ref{sec:intro}.

Some studies use solvers more directly to construct test suites for SUTs with constraints.  
In these studies, the problem of finding a test suite of a given size is 
transformed into a combinatorial problem and then the combinatorial problem is solved 
by a solver. 
Examples of these studies include~\citep{DBLP:journals/jar/CalvagnaG10,DBLP:journals/ieicet/NanbaTK12,DBLP:conf/icst/YamadaKACOB15}. 


There are some studies that use BDDs and their variants, such as 
\emph{Multi-Valued Decision Diagrams} (MDDs)~\citep{MDD} and 
\emph{Zero-suppressed Decision Diagram} (ZDDs)~\citep{Minato:1993:ZBS:157485.164890}, for CIT. 
In~\citep{Salecker:2011:CPI:2004685.2005510} a one-test-at-a-time greedy algorithm 
is proposed that uses BDDs. 
According to the data reported in~\citep{Salecker:2011:CPI:2004685.2005510}, 
the algorithm was able to generate smaller test suites but required larger running time 
than PICT. 
A BDD is used to represent the set of valid test cases in the same way as described in Section~\ref{subsec:ordinary}, 
except that $|D_i|$, instead of $\log_2 |D_i|$, Boolean variables are used for each parameter $P_i$. 

In~\citep{Segall:2011:UBD:2001420.2001451}, a 
one-test-at-a-time algorithm based on BDDs is proposed. 
The algorithm also represents the space of valid test cases as presented in Section~\ref{subsec:ordinary} 
but uses BDDs not only for validity checking but also for generating  
locally optimal or suboptimal test cases. 
Specifically, set operations based on BDDs are extensively used to select a valid test case that covers many 
interactions yet to be covered. 
This algorithm is adopted by the IBM FOCUS tool which has been in use in industry~\citep{Tzoref-Brill:2018:MES:3236024.3236067}. 
Since this tool is not publicly available and the purpose of our paper is 
to evaluate the performance of constraint handling, we did not 
compare its performance with the approaches we considered in this paper. 

In~\citep{Gargantini2014corrected}, MDDs are used to represent the space of valid test cases. 
In~\citep{DBLP:conf/prdc/OhashiT17}, ZDDs are used to handle a large number of 
interactions that occur in the course of constructing high strength CIT test suites. 
Constraints are not considered in~\citep{DBLP:conf/prdc/OhashiT17}.

\section{Threats to validity}\label{sec:threats} 

One potential threat concerns the representativeness of the problem instances used in 
the experiments. 
We used a total of 62 problem instances but they only represent a part of 
the reality of CIT. 
In addition, some of them are purely synthetic and thus may not 
necessarily capture features that occur in actual software. 
Another potential threat concerns the representativeness of ACTS, which was chosen for comparison 
with the proposed techniques. 
As stated in Section~\ref{sec:evaluation}, we chose ACTS for comparison for some good reasons. 
However, there are other test case generation tools that employ different algorithms. 
This potential threat could be mitigated by incorporating the proposed constraint handling techniques  
in a tool other than ACTS (for example, PICT~\citep{Czerwonka06}) and by conducting performance evaluation. 
We leave it for future work. 


\section{Conclusions}\label{sec:conclusion}

In this paper, we investigated the usefulness of BDD-based constraint handling approaches for CIT.  
The main task of constraint handling is to perform a validity check, that is, 
determine if a full or partial test case is valid or not. 
We considered two types of BDD-based approaches. 
The first one is straightforward one, while the other uses a new technique 
which allows a single BDD to represent a set of all valid full and partial test cases. 
With the BDD, one can perform a validity check simply by traversing the BDD from the root node. 
We developed a program that implements IPOG, an existing test case generating algorithm, 
together with the BDD-based constraint handling approaches and empirically compared its performance 
with ACTS, a state-of-the-art tool which implements IPOG. 
In the experiments conducted, the new BDD-based approach exhibited best performance 
with respect to the time required for test case generation. 

For future work, we plan to incorporate the BDD-based IPOG algorithm into 
our test generation tool CIT-BACH which currently only implements a one-row-at-a-time greedy algorithm. 
Experiments using more problem instances are also worth conducting.

\begin{acknowledgements}
We are grateful for Kazuhiro Ishihara at VALTES Co., Ltd. for research collaboration, including 
provision of an SUT model. 

\end{acknowledgements}

%
%




\end{document}